\documentclass[11pt]{article}
\usepackage{amsmath}
\usepackage{amssymb}
\usepackage{amsthm,amsxtra}
\usepackage{epsf}
\usepackage{eepic}
\newlength{\mytopmargin}
\newlength{\myleftmargin}
\newtheorem{thm}{Theorem}
\newtheorem{cor}{Corollary}
\newtheorem{lemma}{Lemma}
\newtheorem{prop}{Proposition}
\newcommand{\zz}{\mathbb Z}
\newcommand{\qq}{\mathbb Q}

\newcommand\psymmU{%
\begin{picture}(1,1)(0,0)%
\allinethickness{0.5pt}%
\path(0,0)(0,1)(1,1)(1,0)(0,0)%
\end{picture}}
\newcommand\psymmUU{%
\begin{picture}(1,1)(0,0)%
\allinethickness{0.5pt}%
\path(0,0)(0,1)(1,1)(1,0)(0,0)%
\put(0.5,0.5){\makebox(0,0){$\cdot$}}%
\end{picture}}
\newcommand\psymmO{%
\begin{picture}(1,1)(0,0)%
\allinethickness{0.5pt}%
\path(0,0)(0,1)(1,1)(1,0)(0,0)%
\path(0,0)(1,1)%
\end{picture}}
\newcommand\psymmS{%
\begin{picture}(1,1)(0,0)%
\allinethickness{0.5pt}%
\path(0,0)(0,1)(1,1)(1,0)(0,0)%
\path(1,0)(0,1)%
\end{picture}}
\newcommand\psymmu{%
\begin{picture}(1,1)(0,0)%
\allinethickness{0.5pt}%
\path(0,0)(0,1)(1,1)(1,0)(0,0)%
\path(0,0)(1,1)%
\path(0,1)(1,0)%
\end{picture}}

\newbox\tsymmUbox
\newbox\tsymmUUbox
\newbox\tsymmObox
\newbox\tsymmSbox
\newbox\tsymmubox
\setbox\tsymmUbox =\hbox{\kern0.75pt\setlength{\unitlength}{6pt}\psymmU \kern0.75pt}

\setbox\tsymmUUbox=\hbox{\kern0.75pt\setlength{\unitlength}{6pt}\psymmUU\kern0.75pt}
\setbox\tsymmObox =\hbox{\kern0.75pt\setlength{\unitlength}{6pt}\psymmO \kern0.75pt}
\setbox\tsymmSbox =\hbox{\kern0.75pt\setlength{\unitlength}{6pt}\psymmS \kern0.75pt}
\setbox\tsymmubox =\hbox{\kern0.75pt\setlength{\unitlength}{6pt}\psymmu \kern0.75pt}

\newbox\symmUbox
\newbox\symmUUbox
\newbox\symmObox
\newbox\symmSbox
\newbox\symmubox
\setbox\symmUbox =\hbox{\kern0.75pt\setlength{\unitlength}{4.5pt}\psymmU \kern0.75pt}
\setbox\symmUUbox=\hbox{\kern0.75pt\setlength{\unitlength}{4.5pt}\psymmUU\kern0.75pt}
\setbox\symmObox =\hbox{\kern0.75pt\setlength{\unitlength}{4.5pt}\psymmO \kern0.75pt}
\setbox\symmSbox =\hbox{\kern0.75pt\setlength{\unitlength}{4.5pt}\psymmS \kern0.75pt}
\setbox\symmubox =\hbox{\kern0.75pt\setlength{\unitlength}{4.5pt}\psymmu \kern0.75pt}

\def\symmO{{\copy\symmObox}}
\def\symmS{{\copy\symmSbox}}

\begin{document}

\vspace{4cm}
\noindent
{\bf Increasing subsequences and the hard-to-soft edge transition
in matrix ensembles}

\vspace{5mm}
\noindent
Alexei Borodin${}^*$ and Peter J.~Forrester${}^\dagger$

\noindent
${}^*$School of Mathematics, Institute of Advanced Study, Einstein Drive,
Princeton \\ NJ 08540, USA; ${}^\dagger$
Department of Mathematics and Statistics,
University of Melbourne, \\
Victoria 3010, Australia ;
email: borodine@math.upenn.edu;
p.forrester@ms.unimelb.edu.au

\small
\begin{quote}
Our interest is in the cumulative probabilities Pr$(L(t) \le l)$ for the
maximum length of  increasing subsequences in  Poissonized ensembles
of random permutations, random fixed point free involutions and reversed
random fixed point free involutions. It is shown that these probabilities
are equal to the hard edge gap probability for matrix ensembles with unitary,
orthogonal and symplectic symmetry respectively. The gap probabilities
can be written as a sum over correlations for certain determinantal point
processes. From these expressions a proof can be given that the
limiting form of Pr$(L(t) \le l)$ in the three cases is equal  to the
soft edge gap probability for matrix ensembles with unitary, orthogonal
and symplectic symmetry respectively, thereby reclaiming theorems
due to Baik-Deift-Johansson and Baik-Rains.
\end{quote}

\section{Introduction}
Let $S_N$ denote the set of all permutations of $\{1,2,\dots,N\}$. Let
$\pi \in S_N$ and consider a subsequence of image points
$\{\pi(i_1), \pi(i_2),\dots, \pi(i_k)\}$ where $1 \le i_1 < \cdots <
i_k \le N$. Such a subsequence is referred to as an increasing subsequence
of length $k$ if $\pi(i_1) < \pi(i_2) < \cdots < \pi(i_k)$. For a
given $\pi$, let $L_N(\pi)$ denote the maximum length of all the increasing
subsequences. The question of the distribution of $L_N(\pi) =: L_N$, when
$\pi$ is chosen at random from a uniform distribution on $S_N$, was posed
in the  early 1960's by Ulam. In 1999 the question was answered by
Baik, Deift and Johansson \cite{BDJ99}, who proved
\begin{equation}\label{1.1}
\lim_{N \to \infty} {\rm Pr} \Big (
{L_N - 2 \sqrt{N} \over N^{1/6}} \le s \Big ) = F_2(s),
\end{equation}
where $F_2(s)$ is the scaled cumulative distribution of the largest
eigenvalues for large random Hermitian matrices with complex
Gaussian entries (technically matrices from the Gaussian unitary
ensemble (GUE)) \cite{TW94a}. One should consult \cite{AD99} for a review of
the work on Ulam's problem culminating in the Baik-Deift-Johansson
theorem.

In the course of proving (\ref{1.1}), the exponential generating function of
${\rm Pr}(L_N \le l)$,
\begin{equation}\label{2.1}
e^{-t} D_l(t), \qquad
 D_l(t) := \sum_{N=0}^\infty {t^N \over N!}
{\rm Pr}(L_N \le l),
\end{equation}
was introduced. This quantity itself is the cumulative distribution of a
natural quantity due to Hammersley (see e.g.~\cite{AD95}). Thus consider
the unit square with points chosen at random according to a Poisson
process of rate $t$. Form a continuous piecewise linear path, with positive
slope where defined, connecting $(0,0)$ to $(1,1)$ and only changing slope
at a point. Let $L^{\square}(t)$ denote the length of the longest such
``up/right'' path, where the length is defined as the number of Poisson
points in the path. To see the relation to (\ref{2.1}), label the points
$1,\dots,N$ from left to right, then attach a second label $1,\dots,N$
from bottom to top. In this way each array of $N$ points is associated
with a permutation, and furthermore the fact that the points are chosen 
from a Poisson process implies the uniform disitribution  on the set of
permutations of $N$ symbols.
Up/right paths correspond to increasing subsequences and we
have
\begin{equation}\label{3.0}
{\rm Pr}(L^{\square}(t) \le l) = e^{-t} D_l(t).
\end{equation}
It was proved in \cite{BDJ99} that
\begin{equation}\label{3.1}
\lim_{t \to \infty}
{\rm Pr} \Big ( {L^{\square}(t) - 2 \sqrt{t} \over t^{1/6}} \le
s \Big ) = F_2(s).
\end{equation}
In  fact (\ref{3.1}) suffices to prove (\ref{1.1}), by applying a so
called de-Poissonization lemma \cite{Jo98a}.

Four companion identities to (\ref{3.1}), relating the limiting
distribution of longest paths in certain up/right paths problems to
the limiting distribution of the largest eigenvalue in certain
random matrix ensembles, were found by Baik and Rains \cite{BR01a,
BR01b}. Of these two are independent, in that it was shown that the
other two follow as corollaries
\cite[Theorem 2.5]{BR01a}. For the first, modify the original
longest up/right path problem by requiring that initially only the
region below the line $y=1-x$ of the unit square be filled with Poisson
points of rate $t$; the points above the line are then specified by
the image of the initial points reflected about $y=1-x$. Let
$L^{\symmS}(t)$ refer to the longest up/right path from $(0,0)$ to
$(1,1)$ in this setting. Then one has
\begin{equation}\label{3.2}
\lim_{t \to \infty}
{\rm Pr} \Big ( {L^{\symmS}(t) - 2 \sqrt{t} \over 2^{1/3} t^{1/6}} \le
s \Big ) = F_1(s),
\end{equation}
where $F_1(s)$ is the cumulative distribution of the largest eigenvalue for
large random real symmetric matrices with Gaussian entries
(technically matrices from the Gaussian orthogonal ensemble (GOE)) \cite{TW96}.
The random variable $L^{\symmS}(t)$ is related to the maximum length
$L_{2N}^{\symmS}$ 
of all {\it decreasing} subsequences of random fixed point
free involutions $(\pi^2 = \pi, \: \pi(i) \ne i)$ for any $i$) of
$\{1,2,\dots,2N\}$, or equivalently of all increasing subsequences of
reversed fixed point free involutions. Thus
\begin{equation}\label{4.0}
{\rm Pr}(L^{\symmS}(t) \le l) = e^{-t/2} 
D_l^{\symmS}(t), \qquad
 D_l^{\symmS}(t) := \sum_{N=0}^\infty
{t^N \over 2^N} {{\rm Pr}(L_{2N}^{\symmS} \le l) \over (2N)!}.
\end{equation}

For the second of the companion identities, the original longest
up/right path problem is modified by requiring that initially only the
region below the line $y=x$ of the unit square be filled with Poisson
points at rate $t$, with the points above the diagonal specified as the
image of these points reflected about $y=x$. With $L^{\symmO}(t)$ referring
to the longest up/right path in this setting, one has 
\begin{equation}\label{4.1}
\lim_{t \to \infty}
{\rm Pr} \Big ( {L^{\symmO}(t) - 2 \sqrt{t} \over 2^{1/3}t^{1/6}} \le
s \Big ) = F_4(s)
\end{equation}
where $F_4(s)$ is the scaled cumulative distribution of the largest
eigenvalue for large random Hermitian matrices with real quaternion
elements (technically matrices from the Gaussian symplectic ensemble (GSE)).
With $L_{2N}^{\symmO}$ denoting the maximum length of increasing subsequences
of random fixed point free involutions of $\{1,2,\dots,2N\}$, one has
\begin{equation}\label{5.0}
{\rm Pr}(L^{\symmO}(t) \le l) = e^{-t/2} D_l^{\symmO}(t), \qquad
 D_l^{\symmO}(t) := \sum_{N=0}^\infty
{t^N \over 2^N} {{\rm Pr}(L_{2N}^{\symmO} \le l) \over (2N)!}.
\end{equation}

In this paper we will give new proofs of the results (\ref{3.1}),
(\ref{3.2}) and (\ref{4.1}). The original proof of (\ref{3.1}) uses
a Riemann-Hilbert analysis \cite{BDJ99}. The subsequent proofs of
(\ref{3.1}) given in
\cite{Jo00,BOO99} rely on proving the convergence of a certain
Fredholm integral operator determing Pr$(L^\square(t) 
\le l)$ to the Fredholm integral
operator determining $F_2(s)$. 
A combinatorial proof exploiting the interplay between maps and
ramified coverings of the sphere is given in \cite{Ok00}.
In the cases of (\ref{3.2}) and
(\ref{4.1}), a Riemann-Hilbert analysis was again used in the 
original proof \cite{BR01b}. No other derivations of (\ref{3.2})
and (\ref{4.1}) have previously been given. 
Our derivation relies on finding expressions for
Pr$(L^\square(t) 
\le l)$, Pr$(L^{\symmS}(t) \le l)$, 
Pr$(L^\symmO(t) \le l)$ as sums over
correlations determining the probability of the interval $(0,t)$
being eigenvalue free in
the infinite, scaled Laguerre
unitary ensemble (LUE), Laguerre orthogonal ensemble (LOE), and Laguerre
symplectic ensemble (LSE) with parameter value $a=l$. The latter are
known as hard edge gap probabilities.
The probabilities  $F_2(s)$, $F_1(s)$ and
$F_4(s)$, give the so called soft edge gap probability that the
interval $(s,\infty)$ is eigenvalue free in the infinite scaled GUE,
GOE and GSE respectively, and can also be written as a sum over
correlations.
The limit formulas (\ref{3.1}),
(\ref{3.2}) and (\ref{4.1}) are then proved by establishing the convergence
as $a \to \infty$ of the sum over correlations determining the hard
edge gap probabilites, to the sum over correlations determining the
soft edge gap probabilities. Related studies of the convergence of the finite
$N$ {\it soft} edge gap probability in the LOE and LUE
to the corresponding scaled soft edge gap probability
has previously been undertaken in \cite{Jo00a,Jo01,So01}, and it is the
method of \cite{So01} which we adopt here. 

In Section 2 we
present  formulas from the theory of zonal polynomials which allow
Pr$(L^\square(t) 
\le l)$, Pr$(L^{\symmS}(t) \le l)$ and
Pr$L^\symmO(t) \le l)$ to be expressed as hard edge gap probabilities.
In Section 3 we show how the hard edge gap probabilities can be
written as sums over correlations for certain determinantal point
processes, and this exercise is repeated for the soft edge gap
probabilities $F_1(s), F_2(s)$ and $F_4(s)$. The convergence of the
sum over correlations determining the hard edge gap probabilities, to
the sum over correlations determining the soft edge gap probabilities,
is established in Section 4.

\section{Averages over classical groups and the hard edge gap
probability in
the Laguerre ensemble}
\setcounter{equation}{0}
It has been shown by Rains \cite{Ra98} that the generating functions
of interest each can be written as averages over classical groups. Thus
\begin{eqnarray}
D_l(t) & = & \langle e^{\sqrt{t} {\rm Tr}(U+U^\dagger)} \rangle_{U \in
U(l)} \label{6.1} \\
D_l^{\symmS}(t) & = & 
\langle e^{\sqrt{t} {\rm Tr}\,S} \rangle_{S \in  Sp
(l)} \label{6.2} \\
D_{2l}^{\symmO}(t) & = &  \langle e^{\sqrt{t} {\rm Tr}\,O} \rangle_{O \in
O(2l)}  \label{6.3}
\end{eqnarray}
(we use the notation $Sp(l)$ to denote $l \times l$ unitary matrices with
real quaternions elements, or equivalently  $2l \times 2l$ symplectic
unitary matrices with complex elements).

The average (\ref{6.1}) earlier appeared as the cumulative distribution of 
 the smallest eigenvalue for the scaled LUE \cite{Fo93c}. We recall the LUE
refers to the eigenvalue probability density function
\begin{equation}\label{7.1}
{1 \over C} \prod_{l=1}^N \lambda_l^a e^{-\lambda_l}
\prod_{1 \le j < k \le N} (\lambda_k - \lambda_j)^2, \quad
\lambda_l >0.
\end{equation}
For $a=n-N$, $n \ge N$, it is realized by eigenvalues of the matrix
$X^\dagger X$, where $X$ is a $n \times N$ complex Gaussian matrix. 
The cumulative distribution of the smallest eigenvalue for the ensemble
(\ref{7.1}) (or what is the same thing, the probability of no
eigenvalues in the interval $(0,s)$), to be denoted
$E_2^L(s;a;N)$, is obtained from (\ref{7.1}) by integrating each of the
eigenvalues over $(s,\infty)$,
\begin{equation}\label{2.4'}
E_2^L(s;a;N) := {1 \over C} \int_s^\infty d \lambda_1 \,
 \lambda_1^a e^{-\lambda_1} \cdots 
 \int_s^\infty d \lambda_N \,
 \lambda_N^a e^{-\lambda_N} \prod_{1 \le j < k \le N}
(\lambda_k - \lambda_j)^2
\end{equation}
(the normalization $C$ is such that $E_2^L(0;a;N) = 1$).
Rescaling
the eigenvalues in  
the vicinity of the origin (referred to as the hard edge since the
eigenvalues are restricted to $\lambda_l >0$) by
\begin{equation}\label{7.1a}
\lambda_l \mapsto {x_l \over 4N},
\end{equation}
 the correlations have a well defined $N \to \infty$ limit \cite{Fo93a}.
This implies the scaled gap probability
\begin{equation}\label{2.sg}
E^{L \, \rm hard}_2(s;a) := \lim_{N \to \infty}
E_2^L \Big ( {s \over 4N};a;N \Big )
\end{equation}
exists. Moreover, it was shown in \cite{FH94} that for
$a \in \zz_{\ge 0}$, $E_2^L(s;a;N)$ has a simple structure, allowing
it, and its scaled limit, to be expressed as an $a \times a$ determinant.
The determinant can alternatively be written as an $a$-dimensional
integral, giving \cite{Fo93c}
\begin{equation}\label{7.1a'}
E^{L \, \rm hard}_2(s;a) = e^{-s/4} \langle e^{{1 \over 2} \sqrt{s}
{\rm Tr}\, (U + U^\dagger)} \rangle_{U \in U(a)},
\end{equation}
and thus relating to (\ref{6.1}).

The averages (\ref{6.2}) and (\ref{6.3}) have recently been shown to be
equal to the cumulative distribution of the smallest eigenvalue
in the scaled, infinite LOE and
LSE respectively \cite{FW02}. These matrix ensembles refer to the 
eigenvalue probability density function
\begin{equation}\label{7.2}
{1 \over C} \prod_{l=1}^N \lambda_l^a e^{-c\lambda_l/2}
\prod_{1 \le j < k \le N} |\lambda_k - \lambda_j|^\beta, \quad
\lambda_l >0,
\end{equation}
where $\beta=1$, $c=1$ for the LOE, and $\beta=4$, $c=2$ for the LSE. For
$a=(n-N-1)/2$, $n \ge N$, the LOE is realized by random matrices of the form
$X^T X$, where $X$ is a $n \times N$ real standard Gaussian matrix, while
for $a=2(n-N)+1$, matrices of the form $X^\dagger X$ with $X$ a
$n \times N$ real quaternion Gaussian matrix (embedded as a complex
matrix) realizes the LSE (see e.g.~\cite{Fo02}). 
For general parameters $a,c,\beta$ in (\ref{7.2}), analogous to
(\ref{2.4'}) we define the gap probability
\begin{equation}\label{2.7'}
E_\beta^L(s;a,c;N) := {1 \over C} \int_s^\infty d \lambda_1 \,
 \lambda_1^a e^{-c\lambda_1/2} \cdots
 \int_s^\infty d \lambda_N \,
 \lambda_N^a e^{-c\lambda_N/2} \prod_{1 \le j < k \le N}
|\lambda_k - \lambda_j|^\beta.
\end{equation}
As an aside we remark that a random matrix construction of the general
$\beta$ Laguerre ensemble has recently been given \cite{DE02}.
For the scaled limits of the LOE and LSE cases one defines
\begin{eqnarray}
E^{L \, \rm hard}_1(s;a) &:=& \lim_{N \to \infty}
E_1^L \Big ( {s \over 4N};a,1;N \Big ) \label{2.7a}
\\
E^{L \, \rm hard}_4(s;a) &:=& \lim_{N \to \infty}
E_4^L \Big ( {s \over 4N};a,2;N/2 \Big ) \label{2.7b}
\end{eqnarray}
(in (\ref{2.7b}) it is assumed $N$ is even; the use of $N/2$ therein
comes about naturally in studying the inter-relationships between
$E_1^L$, $E_2^L$ and $E_4^L$ \cite{FR02}). We know from \cite{FW02}
that
\begin{eqnarray}
E^{L \,\rm hard}_1(s;a) & = & e^{-s/8} \langle e^{{1 \over 2} \sqrt{s}
{\rm Tr} \, S} \rangle_{S \in Sp(a)} \label{8.1} \\
E^{L \, \rm hard}_4(s;2a) & = & e^{-s/8} \langle e^{{1 \over 2} \sqrt{s}
{\rm Tr} \, O} \rangle_{O \in O(2a)}. \label{8.2}
\end{eqnarray}
The derivation of (\ref{7.1a'}) in \cite{Fo93c} is different from the
derivations of (\ref{8.1}) and (\ref{8.2}) in \cite{FW02}.
A unifying derivation can be given, based on properties of zonal polynomials
and corresponding hypergeometric functions, which we will now present.

The zonal polynomials are the special cases $\alpha = 1/2,1$ and 2 of 
more general polynomials --- the Jack polynomials --- which depends on a
continuous parameter $\alpha$. Let us then revise the definition of these
polynomials. 
Let $\kappa := (\kappa_1,\dots,
\kappa_N)$ denote a partition so that $\kappa_i \ge \kappa_j$
$(i < j)$ and $\kappa_i \in \zz_{\ge 0}$. 
The modulus of a partition is defined by $|\kappa| :=
\sum_{i=1}^N \kappa_i$.
Let $m_\kappa$ denote the
monomial symmetric function corresponding to the partition $\kappa$,
and for
partitions $|\kappa| = |\mu|$ define the dominance partial ordering by the
statement that $\kappa > \mu$ if $\kappa \ne \mu$ and
$\sum_{j=1}^p \kappa_j \ge \sum_{j=1}^p \mu_j$ for each $p=1,\dots,N$.
Introduce the Jack polynomials $P_\kappa^{(1/\alpha)}(z_1,\dots,z_N) =:
P_\kappa^{(1/\alpha)}(z)$ as the unique homogeneous polynomials of degree
$|\kappa|$ with the structure
$$
P_\kappa^{(1/\alpha)}(z) = m_\kappa + \sum_{\mu < \kappa} a_{\kappa \mu}
m_\mu
$$
(the $a_{\kappa \mu}$ are some coefficients in $\qq(\alpha)$) and which
satisfy the orthogonality
$$
\langle P_\kappa^{(1/\alpha)}, P_\rho^{(1/\alpha)} \rangle^{(\alpha)}
\propto \delta_{\kappa, \rho}
$$
where
$$
\langle f, g \rangle^{(\alpha)} :=
\int_{-1/2}^{1/2} dx_1 \cdots \int_{-1/2}^{1/2} dx_N \,
\overline{f(z_1,\dots,z_N)} g(z_1,\dots,z_N)
\prod_{1 \le j < k \le N} | z_k - z_j |^{2 \alpha}, \quad
z_j := e^{2 \pi i \theta_j}.
$$
We remark that when $\alpha = 1$ the Jack polynomials coincides with the
Schur polynomials; also we should point out that there are other ways
to define the Jack polynomials  (see e.g.~\cite{Ma95}).
Let
\begin{equation}\label{dk}
d_\kappa' = \prod_{(i,j) \in \kappa}
\Big ( \alpha (a(i,j) + 1) + l(i,j)  \Big ),
\end{equation}
where the notation $(i,j) \in \kappa$ refers to the diagram of $\kappa$, in
which each part $\kappa_i$ becomes the nodes $(i,j)$,
$1 \le j \le \kappa_i$ on a square lattice labelled as is conventional
for a matrix. The quantity $a(i,j)$ is the so called arm length (the number of
nodes in row $i$ to the right of column $j$), while $l(i,j)$ is the leg
length (number of nodes in column $j$ below row $i$). Define the
renormalized Jack polynomial
\begin{equation}\label{cpc}
C_\kappa^{(\alpha)}(z) := {\alpha^{|\kappa|} | \kappa|! \over d_\kappa'}
P_\kappa^{(\alpha)}(z),
\end{equation}
and introduce the generalized factorial function
$$
[u]_{\kappa}^{(\alpha)} =
\prod_{j=1}^N {\Gamma(u - (j-1)/\alpha + \kappa_j) \over
\Gamma(u - (j-1)/\alpha)}.
$$
Then the generalized hypergeometric function ${}_p^{} F_q^{(\alpha)}$
based on the Jack polynomial (\ref{cpc}) is specified by the series
\begin{equation}\label{cpc1}
{}_p^{} F_q^{(\alpha)}(a_1,\dots,a_p;b_1,\dots,b_q;z)
:= \sum_{|\kappa|} {1 \over |\kappa|!}
{[a_1]_\kappa^{(\alpha)} \cdots [a_p]_\kappa^{(\alpha)} \over
[b_1]_\kappa^{(\alpha)} \cdots [b_q]_\kappa^{(\alpha)} }
C_\kappa^{(\alpha)}(z)
\end{equation}
(when $N=1$ this reduces to the classical definition of ${}_p F_q$).
For future reference,
we draw attention to the confluence property
\begin{equation}\label{2.12a}
\lim_{N \to \infty} {}_2^{} F_1^{(\alpha)}(N+a_1, N+a_2;b;z/N^2) =
{}_0^{} F_1^{(\alpha)}(b;z)
\end{equation}
which is derived by recalling the homogeniety property
$C_\kappa^{(\alpha)}(z/N^2) =
N^{-2|\kappa|} C_\kappa^{(\alpha)}(z)$ and taking the limit term-by-term
(the latter is justified since ${}_2 F_1^{(\alpha)}$ is analytic for
$|z_i| < 1$, $i=1,\dots,N$ \cite{Ka93}).

In the special cases $\alpha =1/2, 1$ and 2 the renormalized Jack polynomials
$C_\kappa^{(\alpha)}(z)$ are zonal polynomials 
for symmetric matrices, complex matrices and real quaternion matrices
respectively (see e.g.~\cite{Ma95}). With $X$ such a matrix one defines
$C_\kappa^{(\alpha)}(X) = C_\kappa^{(\alpha)}(\lambda_1,\dots, \lambda_N)$
where $\lambda_1,\dots, \lambda_N$ are the eigenvalues of $X$
(the eigenvalues of a Hermitian matrix with real quaternion elements
are doubly degenerate --- in this case only the distinct eigenvalues are
included). The zonal polynomials have a number of special properties
not shared by the Jack polynomials in general. In particular, one
has \cite{Ja64, Ma95, Ra95}
\begin{eqnarray}
\langle s_{\lambda}(AO) \rangle_{O \in O(n)} & = &
\left \{ \begin{array}{ll}
\displaystyle
{C_\kappa^{(2)}(A A^T) \over C_\kappa^{(2)}(1^n)}, & 2 \kappa = \lambda \\
0, & {\rm otherwise} \end{array} \right. 
\label{9.1} \\
\langle s_{\lambda}(AU) s_\kappa(U^\dagger A^\dagger) 
\rangle_{U \in U(n)} & = & \delta_{\lambda, \kappa}
{C_\kappa^{(1)}(A A^\dagger) \over C_\kappa^{(1)}(1^n)}
\label{9.2} \\
\langle s_{\lambda}(AS) \rangle_{S \in Sp(n)} & = &
\left \{ \begin{array}{ll} \displaystyle
{C_\kappa^{(1/2)}(A A^\dagger) \over C_\kappa^{(1/2)}(1^n)}, & 
\kappa^2 = \lambda, 
\\ 0, & {\rm otherwise} \end{array} \right.
\label{9.3}
\end{eqnarray}
where in (\ref{9.1}) the partition $2 \kappa$ is the partition obtained
by doubling each part of $\kappa$, while in (\ref{9.3}),
$\kappa^2$ is the partition obtained by repeating each part of
$\kappa$ twice. Also $C_\kappa^{(\alpha)}(1^n) :=
C_\kappa^{(\alpha)}(z_1,\dots,z_n) |_{z_1 = \cdots = z_n = 1}$.
As an aside we note that these zonal polynomial identities have recently
been conjectured to carry
over to the more general $q$-setting 
\cite{Ra01}.

Of interest to us is a corollary of (\ref{9.1})--(\ref{9.3}).
\begin{cor}\label{c1}
We have
\begin{eqnarray}
\langle e^{{\rm Tr}(AU)}
e^{{\rm Tr}(U^\dagger A^\dagger)} \rangle_{U \in U(n)} & = &
{}_0^{} F_1^{(1)}(n;A A^\dagger) \label{10.6} \\
\langle e^{{\rm Tr}(AS)} \rangle_{S \in Sp(n)} & = &
{}_0^{} F_1^{(1/2)}(2n;A A^\dagger) \label{10.5} \\
\langle e^{{\rm Tr}(AO)} \rangle_{O \in O(n)} & = &
{}_0^{} F_1^{(2)}(n/2;A A^T/4). \label{10.4}
\end{eqnarray}
\end{cor}

\noindent Proof. \quad
Now we know that for general $\alpha$ \cite{St89}
\begin{equation}\label{10.1}
C_\kappa^{(\alpha)}(1^n) = |\kappa|!
{\alpha^{2 |\kappa|} [n/\alpha]_\kappa^{(\alpha)} \over
h_\kappa d_\kappa'}
\end{equation}
where $d_\kappa'$ is specified by (\ref{dk}) and
$$
h_\kappa := \prod_{(i,j) \in \kappa} \Big ( \alpha a(i,j) +
l(i,j) + 1 \Big ).
$$
We also know that (see e.g.~\cite{Ma95})
\begin{equation}\label{10.2}
\sum_{\lambda} {s_\lambda(X) \over d_\lambda' |_{\alpha = 1}} =
\exp {\rm Tr} \, X.
\end{equation}
Consider the identities (\ref{9.1}) and (\ref{9.3}). We multiply both
sides by $1/d_\lambda|_{\alpha = 1}$ and use (\ref{10.2}) on the left
hand sides. On the right hand sides we use (\ref{10.1}) and the easily
verified identities
\begin{equation}\label{10.3}
{(h_\kappa d_\kappa') |_{\alpha = 2} \over
d_{2 \kappa} |_{\alpha = 1}} = 1, \qquad
{ 2^{2 |\kappa|} (h_\kappa d_\kappa') |_{\alpha = 1/2} \over
d_{\kappa^2}} = 1
\end{equation}
together with the definition (\ref{cpc1}) to deduce (\ref{10.5}) and
(\ref{10.4}). The identity (\ref{10.6}) results by first multiplying
both sides of (\ref{9.2}) by $1/(d_\lambda |_{\alpha = 1} )^2$,
making use of (\ref{10.1}) (note that for $\alpha = 1$, $d_\kappa' =
h_\kappa$), then using (\ref{10.2}) and (\ref{cpc1}).
\hfill $\square$

We remark that the first two identities of Corollary \ref{c1} are due
to James \cite{Ja64}, while the third, which is implicit in the work of
Rains
\cite{Ra95}, appears not to have appeared in print before.
 
The probability $E_\beta^L(s;a,c;N)$ as specified by (\ref{2.7'}), 
for general $\beta > 0$, general $N$, scale chosen with $c=\beta$ and
$a \in \zz_{\ge 0}$ has been given in \cite{Fo93c} in terms of the
generalized hypergeometric function ${}_1^{} F_1^{(\beta/2)}$. In the
scaled $N \to \infty$ limit this same probability was given in
terms of the generalized hypergeometric function ${}_0^{}
F_1^{(\beta/2)}$. Thus
 one has
\begin{equation}\label{11.1}
{E}_\beta^{L \, \rm hard}(s;a,\beta) 
:= \lim_{N \to \infty} E_\beta^L \Big ( {s \over 4N};a,\beta;N \Big )
= e^{-\beta s /8}
{}_0^{} F_1^{(\beta/2)} \Big ( {2a \over \beta}; x_1,\dots, x_a
\Big ) \Big |_{x_1 = \cdots = x_a = s/4}.
\end{equation}
For  $\beta = 1$ and $\beta = 2$ the definition of 
${E}_\beta^{L \, \rm hard}(s;a,\beta)$ coincides with the definition of
${E}_1^{L \, \rm hard}(s;a)$ given by (\ref{2.7a}) and
${E}_2^{L \, \rm hard}(s;a)$ given by (\ref{2.sg}) respectively, so we have
\begin{eqnarray}
{E}_1^{L \, \rm hard}(s;a) & = & e^{-s/8} {}_0F_1^{(1/2)}(2a;x_1,\dots,
x_a) \Big |_{x_1 = \cdots = x_a = s/4} \label{2.sg1}\\
{E}_2^{L \, \rm hard}(s;a) & = & e^{-s/4} {}_0F_1^{(1)}(a;x_1,\dots,
x_a) \Big |_{x_1 = \cdots = x_a = s/4}. \label{2.sg2}
\end{eqnarray}
However, in the case $\beta = 4$
the choice $c= \beta$ used to derive (\ref{11.1})
doesn't agree with the convention used to specify
$E^{L \, \rm hard}_4(s;a)$ in (\ref{2.7b}), 
which also had the pecularity of first having
$N$ replaced by $N/2$ before the $N \to \infty$ limit is taken. As a
result, we have
\begin{equation}\label{11.2}
E^{L \, \rm hard}_4(s;a) = {E}_\beta^{L \, \rm hard}(s/4;a,c) 
\Big |_{\beta = c = 4}
\end{equation}
and consequently
\begin{equation}\label{2.sg3}
{E}_4^{L \, \rm hard}(s;a)  =  e^{-s/8} {}_0^{}F_1^{(2)}(a/2;x_1,\dots,
x_a) \Big |_{x_1 = \cdots = x_a = s/16}.
\end{equation}

The identities (\ref{7.1a'}), (\ref{8.1}) and (\ref{8.2}) can now be
reclaimed. Thus in Corollary \ref{c1}
we choose $A = (\sqrt{t}/2)  I_n$, $I_n$ denoting the $n \times n$
identity matrix, and substitute the resulting forms
in (\ref{2.sg2}), (\ref{2.sg1}) and (\ref{2.sg3})
respectively.

With this side issue resolved, let us now explicitly state the
implication of the identities (\ref{7.1a'}), 
(\ref{8.1}) and (\ref{8.2}) in relation to the
probabilities Pr$(L^{\square}(t) \le l)$,
Pr$(L^{\symmS}(t) \le l)$ and Pr$(L^{\symmO}(t) \le l)$.

\begin{prop}\label{p1}
We have
\begin{eqnarray*}
{\rm Pr} (L^{\square}(t/4) \le l) & = & E_2^{L \, {\rm hard}}(t;l) \\
{\rm Pr} (L^{\symmS}(t/4) \le l) & = & E_1^{L \, {\rm hard}}(t;l) \\
{\rm Pr} (L^{\symmO}(t/4) \le l) & = & E_4^{L \, {\rm hard}}(t;2l).
\end{eqnarray*}
\end{prop}

\noindent Proof. \quad
Substitute (\ref{7.1a'}), (\ref{8.1}) and (\ref{8.2})
in
the identities (\ref{6.1})--(\ref{6.3}) respectively, and then
substitute
the new from
of (\ref{6.1})--(\ref{6.3}) in (\ref{3.0}), (\ref{3.2}) and
(\ref{4.0}) respectively.
\hfill $\square$

\section{The gap probabilities as a sum over $k$-point correlations}
\setcounter{equation}{0}
We seek formulas for the hard edge gap probability in
Proposition \ref{p1} which enable the scaled $t,l \to \infty$ limit
to be analyzed. For this purpose we make use of the well known (and
simple to derive) fact that for a general eigenvalue probability
density function the probability $E^{(N)}(I)$ of having no
eigenvalues in an interval $I$ is given as a sum over the corresponding
$k$-point correlations, 
\begin{equation}\label{A.2}
E^{(N)}(I) = 1 + \sum_{k=1}^N {(-1)^k \over k!}
\int_Idx_1 \cdots \int_I dx_k \, \rho_k^{(N)}(x_1,\dots,x_k).
\end{equation}
Let $A_N(x) = \alpha_N + a_N x$ define a linear scale such that
for $x_1,\dots,x_k$ fixed
\begin{equation}\label{A.3}
\lim_{N \to \infty} a_N^k  \rho_k^{(N)}(A_N(x_1),\dots,A_N(x_k)) =
 \rho_k(x_1,\dots,x_k)
\end{equation}
where $\rho_k$ denotes the limiting distribution. 
We would like to be able to write
$\lim_{N \to \infty} E^{(N)}(A_N(I))$ as the right hand side of
(\ref{A.2}) with $\rho_k^{(N)}$ replaced by $\rho_k$. Sufficient
conditions for this to hold are given by the following specialization of
a recent lemma due to Soshnikov \cite{So01}.

\begin{prop}\label{ps.1}
Consider a sequence of point processes labelled by $N$. Suppose that after
the linear scaling $x_j \mapsto A_N(x_j)$ of each of the coordinates, the
sequence approaches a limit point process with 
correlations $\{\rho_k \}_{k=1,2,\dots}$ such that
\begin{equation}\label{3.3f}
\sum_{k=1}^\infty \Big ( {1 \over k!} \int_I dx_1 \cdots
\int_I dx_k \, \rho_{k}(x_1,\dots,x_k) \Big )^{-1/k}
\end{equation}
diverges,
 and suppose furthermore that
\begin{equation}\label{A.4a}
\lim_{N \to \infty} a_N^k \int_I dx_1 \cdots \int_I dx_k \,
\rho_k^{(N)}(A_N(x_1),\dots,A_N(x_k)) =
\int_I dx_1 \cdots \int_I dx_k \,
\rho_k(x_1,\dots,x_k).
\end{equation}
Then
\begin{equation}\label{A.4}
E(I)  :=  \lim_{N \to \infty} E^{(N)}(A_N(I)) 
 =  1 + \sum_{k=1}^\infty {(-1)^k \over k!}
\int_I dx_1 \cdots \int_I dx_k \, \rho_{k}(x_1,\dots,x_k).
\end{equation}
\end{prop}
The significance of the condition that (\ref{3.3f}) diverges is
that it implies  \cite{Le73,So00} the limit process to then have
the property that for any $I \subset \mathbb R$ the distribution of the
number of particles in $I$ is uniquely determined by the
correlation functions of the process.
Observe that a sufficient condition for (\ref{3.3f}) to diverge 
is that
\begin{equation}\label{A.5}
\int_I dx_1 \cdots \int_I dx_k \,
\rho_k(x_1,\dots,x_k) = o (k!),
\end{equation}
which itself is required
for the series in (\ref{A.4})
to be convergent. If we assume (\ref{A.5}), then (\ref{A.4}) follows
from (\ref{A.4a}) by dominated convergence.

Let us now compute the explicit form of (\ref{A.4}) in the case of the
LUE hard edge gap probability. The form of $\rho_k^{(N)}$ in this case
has a structure common to all probability density functions of the
form
\begin{equation}\label{D.1}
{1 \over C} \prod_{l=1}^N w_2(\lambda_l)
\prod_{1 \le j < k \le N} (\lambda_k - \lambda_j)^2.
\end{equation}
Thus with $w_2(x)$ in (\ref{D.1}) non-negative but otherwise general the
corresponding
$k$-point correlations are given by 
\begin{equation}\label{3.9}
\rho_{k}^{(N)}(x_1,\dots,x_k) = \det [ K(x_j,x_l) ]_{j,l=1,\dots,k},
\end{equation}
where
\begin{equation}\label{3.10}
K(x,y) := {(w_2(x) w_2(y))^{1/2} \over (p_{N-1}, p_{N-1})_2}
{p_N(x) p_{N-1}(y) - p_N(y) p_{N-1}(x) \over x - y}.
\end{equation}
In (\ref{3.10}) $\{p_j(x) \}_{j=0,1,\dots}$ is the set of monic polynomials
orthogonal with respect to $w_2(x)$, and $(p_n,p_n)_2$ is the
corresponding normalization. The LUE is the special case
$w_2(x) = x^a e^{-x}$ $(x > 0)$ of (\ref{D.1}). Denote (\ref{3.10})
in this case by $K^L$. Then we know \cite{Fo93a}
that for fixed $x,y>0$ 
\begin{equation}\label{4.1a}
\lim_{N \to \infty} {1 \over 4N} K^L\Big ( {x \over 4N},
{y \over 4N} \Big ) = K^{\rm Bessel}_2(x,y),
\end{equation}
\begin{equation}\label{3.41a}
K^{\rm Bessel}(x,y) :=
{ J_{a}(\sqrt{x}) \sqrt{y}
J_a'(\sqrt{y}) - \sqrt{x} J_{a}'( \sqrt{x})
J_a(\sqrt{y}) \over 2 (x-y) }.
\end{equation}
Furthermore, it has been proved \cite{Bor99} that the convergence in
(\ref{4.1a}) is uniform for $x,y$ in compact sets on the positive
half line. Because $I=[0,s]$ is compact, it follows immediately
that (\ref{A.4a}) holds with $a_N = 1/4N$. The fact that $\rho_k$
is given by the determinant of a $k \times k$ symmetric 
non-negative matrix, the entries of which are independent of $k$,
implies the bound 
\cite{Jo00}
\begin{equation}\label{4.1b}
\rho_{k}(x_1,\dots, x_k) \le \rho_{1}(x_1) \cdots
\rho_{1}(x_k).
\end{equation}
This in turn implies
$$
\int_I dx_1 \cdots \int_I dx_k \, \rho_{k}(x_1,\dots,x_k) =
O(c^k)
$$
for some $c > 0$, so (\ref{A.5}) holds. Consequently, from
Proposition \ref{ps.1}, the scaled gap probability (\ref{2.sg}) can
be written in the following well known form \cite{TW94b}.

\begin{prop}
Let $K_2^{\rm Bessel}$ be given by (\ref{3.41a}).
We have
\begin{equation}\label{L.1}
E^{L \, {\rm hard}}_2(s;a) = 1 +
\sum_{k=1}^\infty {(-1)^k \over k!}
\int_0^s dx_1 \cdots \int_0^s dx_k \, \det [ K^{\rm Bessel}_2(x_j,x_l)
]_{j,l=1,\dots,k}.
\end{equation}
\end{prop}

In the cases $\beta = 1$ and $\beta = 4$ the limiting correlations are
quaternion determinants, or equivalently Pfaffians \cite{FNH99,Du01}.
This leads to a more complicated analysis than that required for the
case $\beta = 2$. However, at the expense of a minor digression, the
computation of $E_1^{L \, \rm hard}$ as a sum over correlations
can also be posed as a problem involving (scalar) determinants rather than
Pfaffians. In addition, by following this route we will reclaim the
known fact \cite{FR02} that $E_4^{L \, \rm hard}(s;a)$ is simply related
to $E_2^{L \, \rm hard}(s;a_2)$ and $E_1^{L \, \rm hard}(s;a_1)$ for
particular $a_1, a_2$, so no independent analysis of
$E_4^{L \, \rm hard}(s;a)$ is required.

We begin our digression by revising that the general $\beta$ Laguerre ensemble
as specified by (\ref{7.2})
 can be viewed as a limiting case of the
Jacobi ensemble, the latter
specified by the eigenvalue probability density function
\begin{equation}\label{20.1}
{1 \over C} \prod_{l=1}^N \lambda_l^a (1 - \lambda_l)^b
\prod_{1 \le j < k \le N} | \lambda_k - \lambda_j |^\beta, \qquad
0 < \lambda_l < 1.
\end{equation}
For $\beta = 1,2$ and 4, and $a=\beta(n_1-m+1)/2-1$, $b=\beta(n_2
-m+1)/2-1$, this is realized by matrices of the form $A(A+B)^{-1}$,
where $A = a^\dagger a$, $B = b^\dagger b$, with $a$, $b$ real
($\beta = 1$), complex $(\beta=2)$ and real quaternion $(\beta = 4)$
Gaussian random matrices of dimension $n_1\times m$, $n_2\times m$
\cite{Mu82,Fo02}.
To obtain (\ref{7.2}) 
from (\ref{20.1}), make the replacement
$\lambda_l \mapsto c \lambda_l /2b$ in (\ref{20.1}) and take the limit
$b \to \infty$. In the vicinity of $\lambda=0$ both (\ref{7.2})
 and
(\ref{20.1}) have the same functional form, and so it is to be anticipated
that after appropriate scaling the local statistical properties will
also be the same. In the case of the hard edge gap probability, this
can readily be demonstrated, as we will now show.

Let $E_\beta^{J}(s;a,b;N)$ denote the probability that there are
no eigenvalues in the interval $(0,s)$ of the Jacobi ensemble
(\ref{20.1}). Then by definition
\begin{eqnarray}
E_\beta^{J}(s;a,b;N) 
&= &{1 \over C} \int_s^1 d \lambda_1 \, \lambda_1^a (1 - \lambda_1)^b
\cdots \int_s^1  d \lambda_N  \, \lambda_N^a (1 - \lambda_N)^b
\prod_{1 \le j < k \le N} | \lambda_k - \lambda_j|^\beta,
\nonumber \\
&=&{1 \over C}(1-s)^{(1+a+b)N + \beta N(N-1)/2} 
\int_0^1 d \lambda_1 \, (\lambda_1 + s/(1-s) )^a(1 - \lambda_1)^b
 \nonumber \\ && 
\cdots \int_0^1 d \lambda_N \, (\lambda_N + s/(1-s) )^a(1 - \lambda_N)^b 
\prod_{1 \le j < k \le N} | \lambda_k - \lambda_j|^\beta.
\nonumber \\
\end{eqnarray}
Notice that for positive integer values of $a$ the multidimensional integral
in this last expression is a polynomial in $s/(1-s)$. From the work of
Kaneko \cite{Ka93} we know that this multidimensional integral can be
written as an $a$-dimensional generalized hypergeometric function
${}_2^{} F_1^{(\beta/2)}$, thus giving
\begin{eqnarray}
E_\beta^{J}(s;a,b;N) 
&= &(1-s)^{(1+a+b)N + \beta N(N-1)/2} \nonumber \\ &&\times
{}_2^{} F_1^{(\beta/2)}\Big (-N;{2 \over \beta}(a+b+1)+N-1,
{2 \over \beta}a;s_1,\dots,s_a \Big ) \Big |_{s_1=\cdots=s_a=-s/(1-s)}
\end{eqnarray}
(the normalization is fixed by requiring that both sides equal unity for
$s=0$). The scaled $N \to \infty$ limit can now be read off using 
(\ref{2.12a}).

\begin{prop}
For $a \in \zz_{\ge 0}$,
\begin{equation}
E_\beta^{J\, \rm hard}(s;a)  := 
\lim_{N \to \infty} E_\beta^{J\, \rm hard}
\Big ( {s \over 4 N^2};a,b;N \Big ) 
 =  e^{-\beta s / 8}
{}_0^{} F_1^{(\beta/2)}\Big ( {2a \over \beta};s_1,\dots,s_a \Big )
 \Big |_{s_1=\cdots=s_a=s/4}.
\end{equation}
\end{prop}
Comparing with (\ref{11.1}) we see that
\begin{equation}\label{11.j}
E_\beta^{J\, \rm hard}(s;a) = {E}_\beta^{L \, \rm hard}(s;a,\beta)
\end{equation}
as anticipated. Thus,  we can drop the superscripts $J$ and $L$ and
simply write $E_\beta^{\rm hard}(s;a)$, where it is to be understood that
in the case $\beta = 4$ we refer to the scaling (\ref{2.7b}). 

There is an advantage in working with the Jacobi ensemble rather than the
Laguerre ensemble. This comes about because of special features of the
case $b=0$ of the former. One such special feature is the formula
\cite{FR02}
\begin{equation}\label{csb} 
E_4^{J}(s/2;a+1,0;N/2) = {1 \over 2} \Big (
E_1^{J}(s;(a-1)/2,0;N) + {E_2^{J}(s;a,0;N) \over
E_1^{J}(s;(a-1)/2,0;N)} \Big ).
\end{equation}
Taking the scaled limit on both sides shows \cite{FR02}
\begin{equation}\label{3.7}
E_4^{\rm hard}(s;a+1) =
{1 \over 2} \Big ( E_1^{\rm hard}(s;(a-1)/2) +
{E_2^{\rm hard}(s;a) \over  E_1^{\rm hard}(s;(a-1)/2)} \Big ),
\end{equation}
thus reducing the study of the $\beta = 4$ case down to that of the
$\beta=1$ and 2 cases. We will see that the analysis of
$E_1^{\rm hard}$ is also made easier by considering the Jacobi ensemble
with $b=0$.

The simplified analysis of $E_1^{\rm hard}$ from this perspective
comes about because
the ensemble JOE$|_{b=0} \cup$ JOE$|_{b=0} =: J^2$, 
formed out of two independent
copies of the JOE with $b=0$, has a simple determinant  form for the
$k$-point distribution of the odd labelled coordinates
(with the eigenvalues ordered $0 < x_1 < x_2 < \cdots < x_{2N} < 1$)
\cite{FR02},
\begin{equation}\label{21.1} 
\rho^{(N) \rm odd}_{(k)}(x_1,\dots,x_k) \Big |_{a \mapsto (a-1)/2} =
\det [ K^{J^2}(x_j, x_l) ]_{j,l=1,\dots,k},
\end{equation}
\begin{equation}\label{21.2}
K^{J^2}(x, y) := - {\partial \over \partial x}
(1 - x)^{1/2} \int_0^{y} (1 - u)^{-1/2} K^J(x,u) \, du
\end{equation}
where $K^J$ is the function (\ref{3.10}) in the Jacobi case with $b=0$ and
the $a$ parameter left unchanged.

To make use of (\ref{21.1}) we follow \cite{Fo99b} and first note
\begin{equation}\label{6.k}
\Big ( E_1^{J}(s;a,0,N) \Big )^2
= E^{J^2}(s;a,0,N) = E^{{\rm odd}(J^2)}(s;a,0,N),
\end{equation}
where the first equality follows from the definition of the ensemble
$J^2$, while the second equality follows from the fact that the
eigenvalues have been labelled so that the first is closest to the
edge at $x=0$. Hence from (\ref{A.2}) and (\ref{21.1})
\begin{equation}\label{207}
\Big ( E_1^{J}(s;(a-1)/2,0,N) \Big )^2 = 1 +
\sum_{k=1}^N {(-1)^k \over k!}
\int_0^s dx_1 \cdots \int_0^s dx_k \,
\det [ K^{J^2}(x_j,x_l)]_{j,l=1,\dots,k}. 
\end{equation}
To proceed further we require the scaled limit of
$K^{J^2}(x,y)$. 

\begin{prop}\label{pa.1}
We have
\begin{equation}\label{k1b}
\lim_{N \to \infty} {1 \over 4 N} K^{J^2}\Big ( {x \over 4 N},
{y \over 4 N} \Big )  = \sqrt{y \over x} 
K_2^{\rm Bessel}(x,y) + {J_a (\sqrt{x}) \over
2 \sqrt{x}} \Big ( 1 -
\int_0^{\sqrt{y}} J_a(t) \, dt \Big ) 
 =:   K_1^{\rm Bessel}(x,y)
\end{equation}
where the convergence is uniform for $x,y \in (0,s)$.
\end{prop}

\noindent
Proof. \quad Using standard uniform estimates for the Jacobi polynomials,
it has been shown in \cite{FR02} that the left hand side of
(\ref{k1b}) converges uniformly for $x,y \in (0,s)$ to
$$
{\partial \over \partial x}
\Big ( x^{1/2} \int_y^\infty v^{-1/2} K^{\rm Bessel}_2(x,v) \, dv
\Big ).
$$
In the same reference, an identity equivalent to the equality between
this expression and $K_1^{\rm Bessel}$ has been given.
\hfill $\square$

It follows from Proposition \ref{pa.1} and (\ref{21.1}) that in addition
to the pointwise convergence of the correlations, the stronger convergence
property (\ref{A.4a}) also holds. Furthermore, noting that $K_1^{\rm 
Bessel}$
is bounded for $x,y \in [0,s]$
(by $M$ say), and Hadamards lemma on bounds for determinants
\cite{WW65} implies that
\begin{equation}\label{8.y}
\rho_k(x_1,\dots,x_k) \le k^{k/2} M^k
\end{equation}
(note that the bound (\ref{4.1b}) no longer necessarily holds because
$K_1^{\rm Bessel}$ is not symmetric). The inequality (\ref{8.y}),
although a gross overestimate of the physically plausible
$\rho_k \le C^k$, is still
sufficient to establish (\ref{A.5}), so we have that both criteria
sufficient for the validity of (\ref{A.4}) hold. Hence the scaled
gap probability can be expanded in the following form,
known but not rigorously justified in \cite{Fo99b}.

\begin{prop}\label{p8a}
We have 
\begin{equation}\label{L.2}
\Big ( E_1^{\rm hard}(s;(a-1)/2) \Big )^2 =
1 + \sum_{k=1}^\infty {(-1)^k \over k!}
\int_0^s dx_1 \cdots \int_0^s dx_k \,
\det [ K_1^{\rm Bessel}(x_j,x_l)]_{j,l=1,\dots,k}
\end{equation}
where $K_1^{\rm Bessel}(x,y)$ is given by (\ref{k1b}).
\end{prop}

Our final preliminary task is to present formulas analogous to
(\ref{L.1}), (\ref{3.7}) and (\ref{L.2}) for the scaled probabilities
$F_2(s)$, $F_1(s)$ and $F_4(s)$ occuring on the right hand side of
(\ref{1.1}), (\ref{3.2}) and (\ref{4.1}) respectively. By definition
$F_1(s)$, $F_2(s)$ and $F_4(s)$ are the scaled probability of no
eigenvalues at the edge of the spectrum of the Gaussian $\beta$-ensemble,
with $\beta = 1,2$ and 4 respectively, the latter being specified by
the eigenvalue probability density function
\begin{equation}\label{G.1}
{1 \over C} \prod_{l=1}^N e^{-c x_l^2/2} \prod_{1 \le j < k \le N}
|x_j - x_k|^\beta.
\end{equation}
Explicitly define
$$
E_\beta^G(s; \beta, c;N) = {1 \over C}
\int_{-\infty}^s dx_1 \cdots \int_{-\infty}^s dx_N \,
\prod_{l=1}^N e^{-c x_l^2/2} \prod_{1 \le j < k \le N}
|x_k - x_j |^\beta.
$$
Then
\begin{eqnarray}\label{3.58'}
F_1(s) &:= & \lim_{N \to \infty} E_1^{G}(\sqrt{2N} + s/\sqrt{2} N^{1/6};
1,1;N) \nonumber \\
F_2(s) &:= & \lim_{N \to \infty} E_2^{G}(\sqrt{2N} + s/\sqrt{2} N^{1/6};
2,2;N) \nonumber \\
F_4(s) &:= & \lim_{N \to \infty} E_2^{G}(\sqrt{2N} + s/\sqrt{2} N^{1/6};
4,2;N/2).
\end{eqnarray}

In general, choosing $(c,N) \mapsto
 (1,N), (2,N), (2,N/2)$ for $\beta = 1,2$ and 4
in (\ref{G.1}) and then scaling the coordinates by
\begin{equation}\label{9.1f}
x_l \mapsto \sqrt{2N} + {x_l \over \sqrt{2} N^{1/6} }
\end{equation}
gives the so called `soft edge' process with parameter $\beta$. The limiting
$k$-point correlation functions have been explicitly computed as
a $k \times k$  determinant in the case $\beta = 2$ \cite{Fo93a},
and a $k \times k$ quaternion determinant (or equivalently Pfaffian) in the
cases $\beta = 1$ and 4 \cite{FNH99}. Moreover, the uniform asymptotic
expansion of the Hermite polynomials \cite{Ol74}
\begin{equation}\label{3.29'}
e^{-x^2/2} H_N(x) = \pi^{-3/4} 2^{N/2 + 1/4} (N!)^{-1/12}
\{ \pi {\rm Ai} (t) + O(e^{-t}) O(N^{-2/3}) \}
\end{equation}
where $x = (2N)^{1/2} + t/2^{1/2} N^{1/6}$, $t \in [t_0, \infty)$,
shows that the correlation functions converge not only pointwise, but also
in the sense of (\ref{A.4a}). In particular, in the case $\beta = 2$
this convergence, together with the bound (\ref{4.1b}) implies
the well known formula \cite{TW94a}
\begin{equation}\label{Ar.1}
F_2(s)  =  
1 + \sum_{k=1}^\infty {(-1)^k \over k!}
\int_s^\infty dx_1 \cdots \int_s^\infty dx_k \,
\det [ K_2^{\rm Airy}(x_j,x_l)]_{j,l=1,\dots,k}
\end{equation}
where
\begin{equation}
 K_2^{\rm Airy}(x,y) =
{{\rm Ai}(x) {\rm Ai}'(y) - {\rm Ai}(y) {\rm Ai}'(x) \over x - y}.
\end{equation}

In the cases $\beta = 1$ and $\beta = 4$ this line of reasoning gives
an expansion for $F_\beta(s)$ of the form (\ref{Ar.1}), but involving
a quaternion determinant in place of the scalar determinant in the case
$\beta = 2$. However, analogous to (\ref{L.2}), in the case
$\beta = 1$ an alternative expansion involving a scalar determinant
can be derived \cite{Fo99b},
\begin{equation}\label{Ar.2}
(F_1(s))^2  =  
1 + \sum_{k=1}^\infty {(-1)^k \over k!}
\int_s^\infty dx_1 \cdots \int_s^\infty dx_k \,
 \det [ K_1^{\rm Airy}(x_j,x_l)]_{j,l=1,\dots,k},
\end{equation}
\begin{equation}\label{k1A}
 K_1^{\rm Airy}(x,y) = K_2^{\rm Airy}(x,y) + {\rm Ai}(x) \Big (
1 - \int_y^\infty {\rm Ai}(v) \, dv \Big ). 
\end{equation}
The workings in \cite{Fo99b} leading to (\ref{Ar.2}) are formal rather
than rigorous. Nonetheless, in the spirit of the chain of argument
leading to (\ref{L.2}), a rigorous derivation of  (\ref{Ar.2}) can
be given.

We recall that in the rigorous derivation of (\ref{L.2}) presented
above, instead of working with the Laguerre ensemble, a particular
Jacobi ensemble was analyzed. This was permitted because it could
be established that the limiting hard edge probability is the same
in both the Laguerre and Jacobi ensemble. 
Likewise we undertake the task of a rigorous derivation
of (\ref{Ar.2}) by analyzing not the finite $N$ GOE, but rather the
finite $N$ LOE with $a=0$. The asymptotic analysis of \cite{FR02}
shows that with the linear change of scale in this latter ensemble,
\begin{equation}\label{9.3e}
x_l \mapsto 4N + 2(2N)^{1/3} x_l,
\end{equation}
the correlation functions converge to the limiting GOE soft edge
correlations in the sense of (\ref{A.4a}). Thus from
Proposition \ref{ps.1} we can regard
$F_1(s)$ as the scaled limit of $E_1^L((s,\infty);0,N)$ where the
latter denotes the probability that there are no eigenvalues in the
interval $(s,\infty)$ of the finite $N$ LOE with $a=0$. The use
of this perspective, analogous to the use of  working with the
$b=0$ Jacobi ensemble rather than the Laguerre ensemble at the
hard edge, is that the ensemble LOE$|_{a=0} \cup$LOE$|_{a=0} =: L^2$,
formed out of two independent copies of the LOE with $a=0$,
has a simple determinant form for the $k$-point correlation of
the {\it even} labelled coordinates (with eigenvalues ordered
$0 < x_1 < x_2 < \cdots < x_{2N} < \infty$) \cite{FR02},
\begin{equation}\label{3.16a}
\rho^{(N) \, \rm even}_{(k)}(x_1,\dots,x_k)  =
\det [ K^{L^2}(x_j, x_l) ]_{j,l=1,\dots,k},
\end{equation}
$$
K^{L^2}(x, y) := - {\partial \over \partial x}
 \int_0^{y}  K^L(x,u) \, du
$$
where $K^L$ is the function (\ref{3.10}) in the Laguerre case with
$a=0$. To make use of (\ref{3.16a}), we note
\begin{equation}\label{6.ka}
\Big ( E_1^{L}((s,\infty);0,N) \Big )^2
= E^{L^2}((s,\infty);0,N) = E^{{\rm even}(L^2)}((s,\infty);0,N)
\end{equation} 
(c.f.~(\ref{6.k})). Now we seek the scaled limit of $K^{L^2}$ and thus the
scaled limit of $\rho_k^{\rm even}$.

\begin{prop}\label{pb.1}
Let $K_1^{\rm Airy}(x,y)$ be given by (\ref{k1A}). We have
\begin{equation}\label{9.4x}
\lim_{N \to \infty} 2(2N)^{1/3} K^{L^2}(4N+ 2(2N)^{1/3} x,
4N+ 2(2N)^{1/3} y) = K_1^{\rm Airy}(x,y) 
\end{equation}
where the convergence is uniform on $x,y \in (s,\infty)$ with remainder
terms which decay exponentially fast in $x$.
\end{prop}

\noindent Proof. \quad This result is essentially contained in \cite{FR02}.
Thus using the uniform asymptotic expansion
\begin{equation}\label{3.38'}
e^{-x/2} x^{a/2} L_n^a(x) = n^{a/2} \Big (
{(-1)^n \over 2^a (2n)^{1/3}} {\rm Ai}(t) + o(n^{-1/3}) O(e^{-t}) \Big )
\end{equation}
where $x=4n+2+2(2n)^{1/3}t$, $t \in [t_0,\infty)$ (c.f.~(\ref{3.29'})),
in  \cite{FR02}
uniform convergence to
\begin{equation}\label{9.5x}
- {\partial \over \partial x} \int_{-\infty}^y K_2^{\rm Airy}(x,t) \, dt
\end{equation}
is established, as is the equality between (\ref{9.5x}) and $K_1^{\rm Airy}$.
The uniform exponentially decaying bound on the error in (\ref{3.38'})
can be used to deduce that the remainder terms in the convergence of
(\ref{9.4x}) decay exponentially fast in $x$. The structure of the
required working is the same as in the proof of Proposition \ref{p.8}
below, so the details will not be presented.
\hfill $\square$

The results of Proposition \ref{pb.1} allow (\ref{Ar.2}) to be established.
Thus, in light of the relation
(\ref{6.ka}), and the evaluation of the scaled limit of 
$\rho_k^{\rm even}$ given by (\ref{3.16a}) and
(\ref{9.4x}) we know from Proposition \ref{ps.1} that  (\ref{Ar.2})
will be valid if the properties (\ref{A.4}) and (\ref{A.5}) can be
verified. Now the structure of the convergence of $K^{L^2}$ in
the scaled limit of  (\ref{9.4x}) substituted in (\ref{3.16a}) shows
immediately that the stronger convergence (\ref{A.4}) holds true.
Furthermore, recalling that ${\rm Ai} (X) = O(e^{-2x^{3/2}/3})$ as
$x \to \infty$ we see that $e^x K_1^{\rm Airy}(x,y)$ is bounded for
$x,y \in [s_0,\infty)$ (by $M$ say) and thus, making use also of
Hadamards lemma we have
$$
\rho_k(x_1,\dots,x_k) \le e^{-(x_1 + \cdots + x_k)} k^{k/2} M^k.
$$
This inequality establishes (\ref{A.5}), thus concluding the working to justify
(\ref{Ar.2}).

It remains to consider the soft edge gap probability for $\beta = 4$.
For this, we have the inter-relationship analogous to (\ref{csb})
\cite{FR02}
\begin{equation}\label{f4f}
F_4(s) = {1 \over 2} \Big ( F_1(s) + {F_2(s) \over F_1(s)} \Big ),
\end{equation}
which was derived as  the limiting form of an exact inter-relationship
between the corresponding finite $N$ gap probabilities in the
Gaussian ensembles.

\section{The hard-to-soft edge transition}
\setcounter{equation}{0}
It is our objective in this, the final section, to prove the following
limit theorem.

\begin{thm}\label{thm1}
For $\beta = 1,2$ let $E_\beta^{\rm hard}(s;a)$ denote the hard edge
gap probability defined by (\ref{2.7a}) and (\ref{2.sg}), and let
$F_\beta(s)$ denote the soft edge gap probability defined by
(\ref{3.58'}). We have
\begin{equation}\label{44.1}
\lim_{a \to \infty} E_\beta^{\rm hard}(a^2 - 2a(a/2)^{1/3}s
+ O(a);ca/2)
= F_\beta(s), 
\end{equation}
where $c=1$ for $\beta = 1$ and $c=2$ for $\beta = 2$.
\end{thm}
Before discussing the proof of this theorem, let us first note an
immediate corollary.

\begin{cor}\label{cv2}
Let $E_4^{\rm hard}(s;a)$ denote the hard edge gap probability defined by
(\ref{2.7b}), and $F_4(s)$ denote the  soft edge gap probability defined in
(\ref{3.58'}). We have that
the limit relation (\ref{44.1}) with $c=2$ holds for $\beta = 4$.
\end{cor}

\noindent
Proof. \quad We substitute $s \mapsto
a^2 - 2a(a/2)^{1/3}s + O(a)$ in (\ref{csb}), and take the limit
$a \to \infty$ on the
right hand side by using (\ref{44.1}). The resulting expression is
precisely the right hand side of (\ref{f4f}). 
\hfill $\square$

Let us return now to Theorem \ref{thm1}. It turns out to be convenient
to prove directly not (\ref{44.1}), but rather the limit theorem
\begin{equation}\label{sh.1}
\lim_{a \to \infty}  E_\beta^{\rm hard}(Q_a(s);ca/2) = F_\beta(s),
\quad Q_a(s) := \Big ( a - \Big ({a \over 2} \Big )^{1/3} s
\Big )^2, \: \: (\beta = 1,2).
\end{equation}
Let us show that if we can prove (\ref{sh.1}) via Proposition
\ref{ps.1}, and thus prove that
\begin{equation}\label{As.2}
\lim_{a \to \infty} \int_0^{Q_a(s)} dx_1 \cdots
\int_0^{Q_a(s)} dx_k \, \rho_k^{\rm hard}(x_1,\dots,x_k) =
\int_s^\infty dx_1 \cdots \int_s^\infty dx_k \,
\rho_k^{\rm soft}(x_1,\dots,x_k)
\end{equation}
where
\begin{eqnarray}
\rho_k^{\rm hard}(x_1,\dots,x_k) & = & \det [K_\beta^{\rm Bessel}
(x_j,x_l) ]_{j,l=1,\dots,k} \label{hb1} \\
\rho_k^{\rm soft}(x_1,\dots,x_k) & = & \det [K_\beta^{\rm Airy}
(x_j,x_l) ]_{j,l=1,\dots,k}
\end{eqnarray}
(the criterium (\ref{A.5}) has already been checked for
$\rho_k^{\rm soft}$ so checking (\ref{As.2}) is sufficient to deduce
(\ref{sh.1})), and further show that
\begin{equation}\label{As.1}
\lim_{a \to \infty} \prod_{l=1}^k (- Q_a'(x_l))
\rho_k^{\rm hard}(Q_a(x_1),\dots,Q_a(x_k)) =
\rho_k^{\rm soft}(x_1,\dots,x_k),
\end{equation}
with convergence uniform on compact sets,
then (\ref{44.1}) holds.

\begin{lemma}\label{le.1}
Define $Q_a(s)$ as in (\ref{sh.1}), and suppose (\ref{As.1}) and (\ref{As.2})
hold. Then
\begin{equation}\label{Cn.1}
\lim_{a \to \infty} \int_0^{A_a(s)+O(a)} dx_1 \cdots
\int_0^{A_a(s)+O(a)} dx_k \, \rho_k^{\rm hard}(x_1,\dots,x_k) =
\int_s^\infty dx_1 \cdots \int_s^\infty dx_k \,
\rho_k^{\rm soft}(x_1,\dots,x_k)
\end{equation}
where $A_a(s) = a^2 - 2a(a/2)^{1/3}s$, and consequently (\ref{44.1}) holds.
\end{lemma}

\noindent
Proof. \quad Now $Q_a(s) = A_a(s) + O(a^{2/3})$, so the left hand side of
(\ref{Cn.1}) is unchanged if we replace $A_a(s)$ by $Q_a(s)$. Doing this, then
noting
\begin{eqnarray*}
&&\lim_{a \to \infty} \int_0^{Q_a(s)+O(a)} dx_1 \cdots
\int_0^{Q_a(s)+O(a)} dx_k \, \rho_k^{\rm hard}(x_1,\dots,x_k) = \\
&& \qquad
\int_{s + O(a^{-1/3})}^{2(a/2)^{2/3}}dx_1 \cdots
\int_{s + O(a^{-1/3})}^{2(a/2)^{2/3}}dx_k \,
\prod_{l=1}^k (- Q_a'(x_l) )
\rho_k^{\rm hard}(Q_a(x_1),\dots,Q_a(x_k))
\end{eqnarray*}
it follows from (\ref{As.2}) that (\ref{Cn.1}) will hold provided
$$
\lim_{a \to \infty} \int_{s + O(a^{-1/3})}^s dx_1 \cdots
\int_{s + O(a^{-1/3})}^s dx_k \,
\prod_{l=1}^k (- Q_a'(x_l) )
\rho_k^{\rm hard}(Q_a(x_1),\dots,Q_a(x_k)) = 0.
$$
But from (\ref{As.1}) the integrand is bounded, so this holds true.
With (\ref{Cn.1}) established, (\ref{44.1}) follows from
Proposition \ref{ps.1}. \hfill $\square$

To establish (\ref{As.1}) and (\ref{As.2}), we first prove the analogue
of Proposition \ref{pb.1}. 

\begin{prop}\label{p.8}
Let $Q_a(s)$ be given as in (\ref{sh.1}). For $\beta = 1,2$, as
$a \to \infty$
\begin{equation}\label{sh.7}
(Q_a'(x) Q_a'(y))^{1/2} K_\beta^{\rm hard}(Q_a(x), Q_a(y))
= K_\beta^{\rm soft}(x,y) + O\left (a^{-1/3} \right ) \left \{
\begin{array}{ll} O(e^{-x}), & \beta = 1 \\ O(e^{-x-y}), & \beta = 2,
\end{array} \right.
\end{equation}
where the remainder terms hold uniformly for $x,y \in (s,\infty)$.
\end{prop}

\noindent
Proof. \quad Consider first the case $\beta = 2$. Recalling
(\ref{3.41a}) and the definition of $Q_a$ from
(\ref{sh.1}), we see asymptotic estimates of
$$
J_a\Big (a - \Big ( {a \over 2} \Big )^{1/3}s \Big ), \qquad
J_a'\Big (a - \Big ( {a \over 2} \Big )^{1/3}s \Big )
$$ 
are required. Now results of Olver \cite[Chapter 11
Sections 10.1--10.4]{Ol74}
imply the uniform asymptotic expansion
\begin{equation}\label{w.1}
J_\nu(\nu z) \sim
{1 \over \nu^{1/3}}
\Big ( {4 \zeta \over 1 - z^2} \Big )^{1/4} \Big \{
{\rm Ai} (\nu^{2/3} \zeta) + O(\nu^{-4/3}) O(e^{-\nu^{2/3} \zeta})
\Big \}
\end{equation}
valid for all $z \in \mathbb C$, arg$\,z \ne \pi$, where
$\zeta = \zeta(z)$ is specified by
$$
{2 \over 3} \zeta^{3/2} = \log {1 + (1-z^2)^{1/2} \over z} -
(1-z^2)^{1/2}.
$$ 
Thus for $z \to 0^+$, $\zeta$ diverges to $+\infty$, then monotonically
decreases to 0 at $z=1$, where it
has the power series expansion 
\begin{equation}\label{w.2}
\zeta(z) = 2^{1/3}(1-z) + O((1-z)^2).
\end{equation}
We remark that
the term $O(e^{-\nu^{2/3} \zeta})$ in (\ref{w.1}) can be
strengthened to involve the exponent $ (\nu^{2/3} \zeta)^{3/2}$, but
this refinement is not needed for our purpose. Let us set 
$z = 1 - 2^{-1/3} w/\nu^{2/3}$, where $0 < w < 2^{1/3} \nu^{2/3}$
and thus  $0 < z < 1$. Making use of (\ref{w.1}), (\ref{w.2})
and the fact that $\zeta(1-z)$ is an increasing function for
$0<z<1$, and using Taylor's theorem to estimate Ai$(w+\epsilon)$,
$|\epsilon| \ll 1$, shows  
\begin{equation}\label{w.3}
J_\nu(\nu-w(\nu/2)^{1/3}) \sim
\Big ( {2 \over \nu} \Big )^{1/3} {\rm Ai}(w) +
O\Big ({1 \over \nu} \Big ) O(e^{-w}).
\end{equation}

To obtain the analogous expansion of $J_\nu'$, we make use of the
formula
\begin{equation}\label{w.4}
J_\nu'(z) = {1 \over 2} (J_{\nu-1}(z) - J_{\nu + 1}(z)).
\end{equation}
Setting $\nu \mapsto \nu \pm 1$, then $z = {\nu \over \nu \pm 1}(1
- w/2^{1/3} \nu^{2/3})$ in (\ref{w.1}) shows
\begin{eqnarray*}
J_{\nu \pm 1}(\nu - w(\nu/2)^{1/3}) & \sim &
\Big ( {2 \over \nu} \Big )^{1/3}
{\rm Ai}\Big (w \pm {2 \over \nu^{1/3}} \Big ) +
O\Big ({1 \over \nu} \Big ) O(e^{-w}) \\
& \sim &
\Big ( {2 \over \nu} \Big )^{1/3}
{\rm Ai}(w ) \pm \Big ( {2 \over \nu} \Big )^{2/3}
{\rm Ai}'(w ) +
O\Big ({1 \over \nu} \Big ) O(e^{-w})
\end{eqnarray*}
where to obtain the second line the Airy function has been expanded
to first order, and the error estimated using Taylor's theorem. 
Substituting this in (\ref{w.4})
shows that apart from the exponent in the term $O(\nu^{-1})$,
(\ref{w.3}) in fact remains valid upon formal
differentiation with respect to $w$, and thus
\begin{equation}\label{w.5}
-\Big ( {\nu \over 2} \Big )^{1/3} J_\nu'(\nu-w(\nu/2)^{1/3}) \sim
\Big ( {2 \over \nu} \Big )^{1/3} {\rm Ai}'(w) +
O(\nu^{-2/3}) O(e^{-w}).
\end{equation}

Making use of (\ref{w.3}) and (\ref{w.5}) in (\ref{3.41a}) with the
substitution $x \mapsto Q_a(x)$, $y \mapsto Q_a(y)$, gives
(\ref{sh.7}) in the case $\beta = 2$, provided $|x-y|$ is
bounded away from zero. This latter proviso is needed at this
stage due to the term $x-y$ in the denominator of the definition
of $K_2^{\rm hard}$, which could affect the decay of the error term.
To see that in fact no such complication arises, for
$|x-y|  \ll 1$, we make use of $K_2^{\rm hard}$
being an analytic function in both $x$ and $y$, and so permitting
the Cauchy-type integral representation 
\begin{equation}\label{w.e}
K_2^{\rm hard}(x,y) = {1 \over 2 \pi} \int_0^{2 \pi}
K_2^{\rm hard}(x,y + R e^{it}) \, dt
\end{equation}
for arbitrary $R > 0$. Choosing $R = 2a(a/2)^{1/3}$ we see that to use
this formula to analyze the left hand side of (\ref{sh.7}) we require
the asymptotic estimates (\ref{w.3}) and (\ref{w.5}) with
$w = y - e^{it} + O(\nu^{-2/3})$. Note that the modulus of the error
terms therein is  $O(e^{-y})$. Substituting in (\ref{w.e}) 
gives the  Cauchy-type
integral representation of $K_2^{\rm soft}(x,y)$ as the leading
term while the remainder term is seen to be bounded by terms
$O(a^{-1/3})O(e^{-x-y})$ coming from the numerator in the definition of
$K_2^{\rm hard}(x,y + R e^{it})$, times the maximum of the scaled denominator
$$
{2a(a/2)^{1/3} \over | Q_a(x) - Q_a(y) + R e^{it}|} \sim
{1 \over | y - x + e^{it} |}.
$$
Because this is bounded for $|x-y| \ll 1$, we see that the error term in
(\ref{w.e}) is indeed as stated in (\ref{sh.7}) for $\beta = 2$.
 
From the definition
(\ref{k1b}), to derive (\ref{sh.7}) in the case $\beta = 1$ the
only remaining task is to give the asymptotic expansion of
$$
\int_0^{Q_a(y)} J_a(t) \, dt = - \int_s^{2(a/2)^{2/3}}
Q_a'(t) J_a(Q_a(t)) \, dt.
$$ 
But this follows immediately from (\ref{w.3}), giving the form required
by (\ref{sh.7}). \hfill $\square$

With the asymptotic formulas (\ref{sh.7}) substituted in (\ref{hb1}), we see
that
$$
\prod_{l=1}^k (- Q_a'(x_l))
\rho_k^{\rm hard}(Q_a(x_1),\dots,Q_a(x_k)) =
\rho_k^{\rm soft}(x_1,\dots,x_k) +
 O\left(a^{-1/3}\right) O(e^{-x_1-\cdots-x_k}).
$$
This shows immediately that (\ref{As.1}) and (\ref{As.2}) hold.
Consequently the limit formula (\ref{sh.1}) is proved, and thus, via
Lemma \ref{le.1}, so is (\ref{44.1}). Now in the limit formula
(\ref{4.1}), extended to the case $\beta = 4$, $c=2$
by Corollary \ref{cv2}, substitute for $E_\beta^{\rm hard}$ according to the
identities of Proposition \ref{p1}. The limit formulas (\ref{3.2}),
(\ref{3.1}) and (\ref{4.1}), equivalent upon de-Poissonization
to limit theorems of Baik, Deift and Johansson
\cite{BDJ99}, and Baik and Rains \cite{BR01b}, are then reclaimed.

\section*{Acknowledgements}
This research was partially conducted during the period AB served
as a Clay Mathematics
Institute Long-Term Prize Fellow, and was supported in part by the
NSF grant DMS-9729992. The work of PJF was supported by the Australian
Research Council. Thanks are due to Eric Rains for pointing out the
significance of the zonal polynomial identities (\ref{9.1})--(\ref{9.3}),
and to Percy Deift for facilitating this collaboration by supporting
a visit of PJF to the University of Pennsylvania during April 2001.


\end{document}